\documentclass[%
reprint,
amsmath,amssymb,
aps,
prl,
]{revtex4-1}

\usepackage{graphicx}
\usepackage{dcolumn}
\usepackage{bm}

\begin{document}


\title{Dirac and Weyl Fermions in \\ $3D$ Hopf-linked Honeycomb Lattices: Hopfene}

\author{Shinichi Saito}
 \email{S.Saito@soton.ac.uk}
\affiliation{Sustainable Electronic Technologies Research Group, Electronics and Computer Science, Faculty of Engineering and Physical Sciences, University of Southampton, SO17 1BJ, UK.}
\author{Isao Tomita}%
\affiliation{%
 Department of Electrical and Computer Engineering, National Institute of Technology, Gifu College, 2236-2 Kamimakuwa, Motosu, Gifu 501-0495, Japan.
}%

\date{\today}

\begin{abstract}
Carbon allotropes such as diamond, nano-tube, Fullerene, and Graphene, have unique lattice symmetries of crystal lattice, but these are topologically trivial.
We have proposed a topologically-nontrivial allotrope, named Hopfene, which has three-dimensional ($3D$) arrays of Hopf-links to bind $2D$ Graphene sheets both vertically and horizontally.
Here, we describe the electronic structures of Hopfene by simple tight-binding calculations.
We confirmed the original Dirac points of $2D$ Graphene were topologically protected upon the introduction of the Hopf links, and low-energy excitations are described by $1D$, $2D$, and $3D$ Dirac and Weyl Fermions.
\end{abstract}

\maketitle

\begin{figure}[h]
\begin{center}
\includegraphics[width=8.6cm]{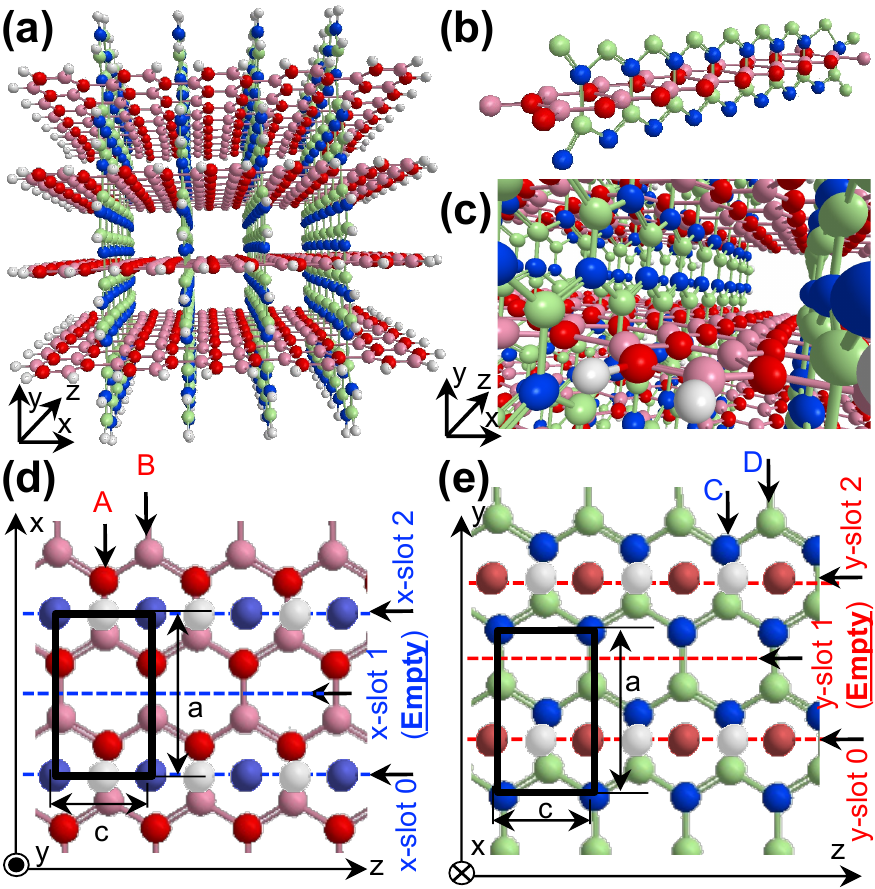}
\caption{
 New $3D$ carbon allotrope, Hopfene.
(a) Crystal structure. 
Graphene sheets are stacked both horizontally and vertically.
Colours of carbon atoms are just guide to the eyes, representing AB- and CD-sublattices.
(b) Hopf-links at the intersection.
(c) Extended view inside Hopfene.
(d) Horizontal and (e) vertical Graphene sheets, showing the penetration of bonds across the sheets perpendicular to each other.
}
\end{center}
\end{figure}

Asymptotic low-energy dispersion of an elementary particle is correlated with the symmetries of the vacuum, and the emergence of the mass is a manifestation of a broken symmetry\cite{Nambu59}.
In condensed-matter physics, we can design materials with certain symmetries, and carbon allotropes are especially useful, because there are huge varieties of families with different translational and rotational symmetries of the crystal lattices in a form of a cage\cite{Kroto85}, a tube\cite{Iijima91,Ando05}, and a sheet\cite{Oshima97,Novoselov04,Ferrari06,Sheng11}.
These materials are revolutionising the material science and technologies, leading to novel applications such as Micro-Electro-Mechanical-Systems (MEMS)\cite{Van18} and Field-Effect-Transistors (FETs)\cite{Novoselov04}, however, these materials are topologically trivial without having a link nor a knot.
On the other hand, there are vast majorities of synthetic materials with non-trivial topological configurations, known in 
polymers\cite{MacGillivray94,Carlucci03,Proserpio10}, macromolecules\cite{Bissell94,Fang09,Molen09,Sauvage17}, and proteins\cite{Dabrowski-Tumanski17}.
Recently, we have applied the concept of topological materials to design a new carbon allotrope, Hopfene, based on a idea of using Hopf-links\cite{Saito19}.

The proposed crystal structure of Hopfene\cite{Saito19} is shown in Fig. 1.
Hopfene is made by stacking two-dimensional ($2D$) honeycomb lattices, Graphene\cite{Oshima97,Novoselov04,Ferrari06} sheets, both vertically and horizontally (Fig. 1 (a)) with arrays of Hopf-links at intersections (Figs. 1 (b) and (c)).
The Graphene sheets are aligned parallel to the direction along zig-zag edges and perpendicular to the arm-chair edges (Figs. 1 (d) and 1 (e)).
In contrast to Carbon-Nano-Tubes (CNTs), for which the structures were categorised by the way to role-up a Graphene sheet, Hopfene is characterised by the way to insert Graphene sheets\cite{Saito19}.
In this example of Fig. 1, available slots for Graphene sheets are occupied every alternative slots, and it is called as (2,2)-Hopfene.
The crystal is tetragonal with the lattice constants of $a=b=3a_0$ within the $x-y$ plane and $c=\sqrt{3} a_0$ along $z$ (Figs. 1 (d) and 1 (e)) at the bond length of $a_0$. 
The details of the Hopfene structures were described in Ref. \cite{Saito19}, and the purpose of this letter is to explore the impacts of Hopf-links on the electronic band diagrams.

Graphene\cite{Oshima97,Novoselov04,Ferrari06} is attracting significant interests, because the low-energy linear excitations are described by massless Dirac Fermions, which are considered to be emerged from magnetic monoples located at valleys of the band structure, called Dirac points\cite{Ando05,Koshino06,Armitage18}.
These monopoles with certain geometrical Pancharatnam-Berry phase factors\cite{Pancharatnam56,Berry84} are topologically protected, so that the electronic structures are robust against the perturbation as far as space and time reversal symmetries are guaranteed\cite{Ando05,Koshino06,Armitage18}.
We have confirmed this feature and found the dimensional crossover of Dirac and Weyl Fermions due to the inter-valley mixing between orthogonal Graphene sheets.

We have employed a tight-binding Hamiltonian for (2,2)-Hopfene\cite{Saito19} (Fig. 1) as 
\begin{eqnarray}
\hat{H}
=
\sum_{{\bf k} \sigma}
\hat{\psi}_{{\bf k} \sigma}^{\dagger}
\begin{pmatrix}
\mathcal{H}_{\rm G}(h_{\rm AB})		& \mathcal{H}_{\rm H} 		 \\
\mathcal{H}_{\rm H}^{\dagger}	& \mathcal{H}_{\rm G}(h_{\rm CD})	 				
\end{pmatrix}
\hat{\psi}_{{\bf k} \sigma},
\end{eqnarray}
where $\hat{\psi}_{{\bf k} \sigma}=\begin{pmatrix} a_{{\bf k} \sigma}, b_{{\bf k} \sigma}, c_{{\bf k} \sigma}, d_{{\bf k} \sigma} \end{pmatrix}$ and $\hat{\psi}_{{\bf k} \sigma}^{\dagger}$ are annihilation and creation operators for electrons with the momentum ${\bf k}$ (in the units of $2\pi(a^{-1},a^{-1},c^{-1})$) and the spin $\sigma$ in A-D sublattices.
The matrix components in Graphene sheets are 
\begin{eqnarray}
\mathcal{H}_{\rm G} (h)
=
\begin{pmatrix}
0				& h 		 \\
h^{*}	& 0 				
\end{pmatrix}
\nonumber
\end{eqnarray}
with $h_{\rm AB}=-t_{\rm G}({\rm e}^{ik_x a/3} +2 {\rm e}^{-ik_x  a/6} \cos (k_z c/2) )$ and $h_{\rm CD}=-t_{\rm G}({\rm e}^{ik_y a/3} +2 {\rm e}^{-ik_y  a/6} \cos (k_z c/2) )$.
The components for Hopf-links are described by a $2\times 2$ matrix
\begin{eqnarray}
\mathcal{H}_{\rm H}
&=
-t_{\rm H}
\left(
\phi_{-\frac{k_x }{6}}^{\dagger}
\phi_{\frac{k_y }{3}}
+ 
\phi_{\frac{k_x }{3}}^{\dagger}
\phi_{-\frac{k_y}{6}}
-2 \cos (\frac{k_z c}{2})
\phi_{-\frac{k_x}{6}}^{\dagger}
\phi_{-\frac{k_y}{6}}
\right), 
\nonumber 
\end{eqnarray}
with $\phi_{k} = \begin{pmatrix} {\rm e}^{ik  a}, {\rm e}^{-ik  a} \end{pmatrix}$. 
The nearest neighbour transfer energies for Graphene and Hopf-links are $t_{\rm G}=2.8$ eV \cite{Ando05,Neto09} and $t_{\rm H}\approx 1.5\pm0.5$ eV, which  was roughly estimated  in comparison with the first principle calculation\cite{Evans14,Tomita19}.
$\hat{H}$ is a $4 \times 4$ matrix with the SU(2) $\otimes $ SU(2) symmetry to describe electron transfers within and between Graphene sheets.
By diagonalising $\hat{H}$ at each ${\bf k}$, we can easily obtain the band diagram with 4 energy dispersions, $E_1 ({\bf k})$-$E_4 ({\bf k})$, measured from the bottom (Fig. 2).

\begin{figure}[t]
\begin{center}
\includegraphics[width=8.6cm]{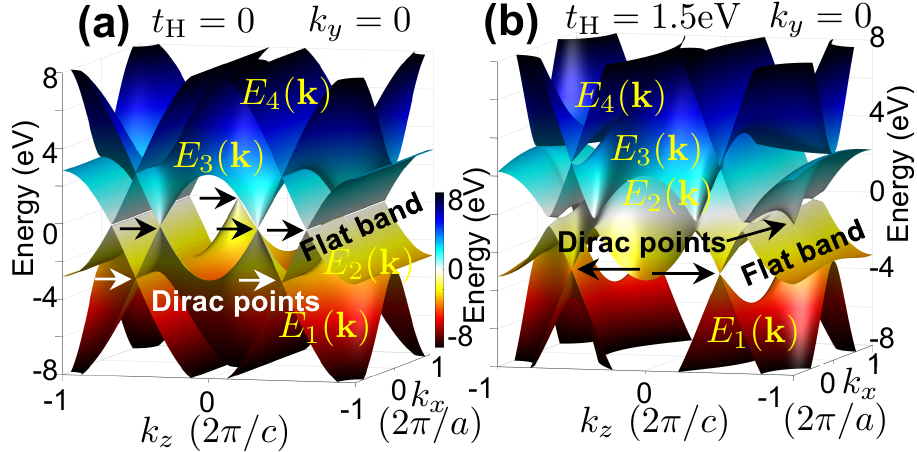}
\caption{
Band structures of (2,2)-Hopfene at $k_y=0$ with the transfer energies of the Hopf-links of (a) $t_{\rm H}=0$ and (b) $t_{\rm H}=1.8$ eV.  
}
\end{center}
\end{figure}

In Graphene, Dirac points are located at $K=(0,-2/3)$ and $K^{'}=(0,-2/3)$, and there are 4 equivalent points at $(\pm 1, \pm 1/3)$ due to the symmetry of the $2D$ honeycomb lattice\cite{Ando05}.
In Hopfene, on the other hand, these 6 Dirac points for horizontal and vertical Graphene sheets are not equivalent, because theses are located in the different planes within the first Brillouin zone. 
Consequently, 10 Dirac points in Hopfene with $t_{\rm H}=0$ are located at $(\pm 1,0,\pm 1/3)$, $(0,\pm 1,\pm 1/3)$, and $(0,0,\pm 2/3)$.
In particular, the last points of  $(0,0,\pm 2/3)$ are degenerate, because these are originated from both horizontal and vertical Graphene sheets.
In the presence of $t_{\rm H}\neq0$, the Dirac points will be shifted in {\bf k}-space, and the slope of the dispersion will be changed, but the essential topological feature will be kept unchanged.

The other interesting feature of the band diagram in Hopfene is the emergence of the flat-bands (Fig. 2).
At $t_{\rm H}=0$, bands will not be dispersive along the direction perpendicular to the Graphene sheets (Fig. 2 (a)), and therefore, the Dirac points will become nodal lines\cite{Armitage18}.
Both Dirac points and nodal lines are topologically protected, so that they cannot disappear by adiabatic turn-on of $t_{\rm H}$ (Fig. 2 (b)), unless Dirac points or nodal lines with opposite geometrical winding numbers would be merged, simultaneously, leading to the destruction of apparent monopoles with opposite magnetic charge.

It was not possible to obtain analytic solution of $\hat{H}$ in general, regardless of its simplicity.
Therefore, we have checked the weak coupling limit, $t_{\rm H} \rightarrow 0$ at $(0,0,2/3)$, where we obtain
\begin{eqnarray}
&\mathcal{H}_{\rm G}(h_{\rm AB})  \approx \hbar v_{\rm G}(\sigma_1 \tilde{k}_z  + \sigma_2 k_x) \nonumber \\
&\mathcal{H}_{\rm G}(h_{\rm CD}) \approx \hbar v_{\rm G}(\sigma_1 \tilde{k}_z  + \sigma_2 k_y) \nonumber \\
& \mathcal{H}_{\rm H}  \approx (-t_{\rm H} + \hbar\tilde{k}_z ){\bf 1}+\frac{2}{3}\hbar v_{\rm H}(i \sigma_3 (k_x-k_y) - \sigma_2 (k_x+ k_y)) \nonumber
\end{eqnarray}
with the velocities of $v_{\rm G}=3t_{\rm G} a_0 /(2\hbar)$ and  $v_{\rm H}=3t_{\rm H} a_0 /(2\hbar)$ for Graphene and Hopf-links, Dirac constant of $\hbar$, $\tilde{k}_z=k_z-2/3$, the identity matrix ${\bf 1}$, and Pauli matrices of $\sigma_1$, $\sigma_2$, and $\sigma_3$.
The constant energy of $-t_{\rm H}$ barely shifts the bands by forming bonding and anti-bonding states between horizontal and vertical Graphene sheets by Hopf-links, and will not change the linearity.
The reset of the Hamiltonian can be diagonalised and we obtained the effective Hamiltonian at $(0,0,2/3)$ as 
\begin{eqnarray}
\hat{H}_{\rm eff}
=
\sum_{{\bf k} \sigma}
\hat{\Psi}_{{\bf k} \sigma}^{\dagger}
\left(
  v_x k_x \sigma_1
+v_y k_y \sigma_2
+v_z k_z \sigma_3
\right)
\hat{\Psi}_{{\bf k} \sigma},
\end{eqnarray}
where $\hat{\Psi}_{{\bf k} \sigma}=\begin{pmatrix} \alpha_{{\bf k} \sigma}, \gamma_{{\bf k} \sigma} \end{pmatrix}$
and $\hat{\Psi}_{{\bf k} \sigma}^{\dagger}$ are annihilation and creation operators of quasi-particles for the Dirac valley and the velocity along $z$ was renormalised to be $v_{z}=(3t_{\rm G}/2 + t_{\rm H}) a_0 /\hbar$, while $v_{x}=v_{y}=v_{\rm G}$.
Therefore, the $3D$ Weyl Fermions will be obtained by the mixing of $2D$ Dirac Fermions at the degenerate Dirac point.
We also obtained the flat-band, which penetrates the Dirac point.

\begin{figure}[t]
\begin{center}
\includegraphics[width=8.6cm]{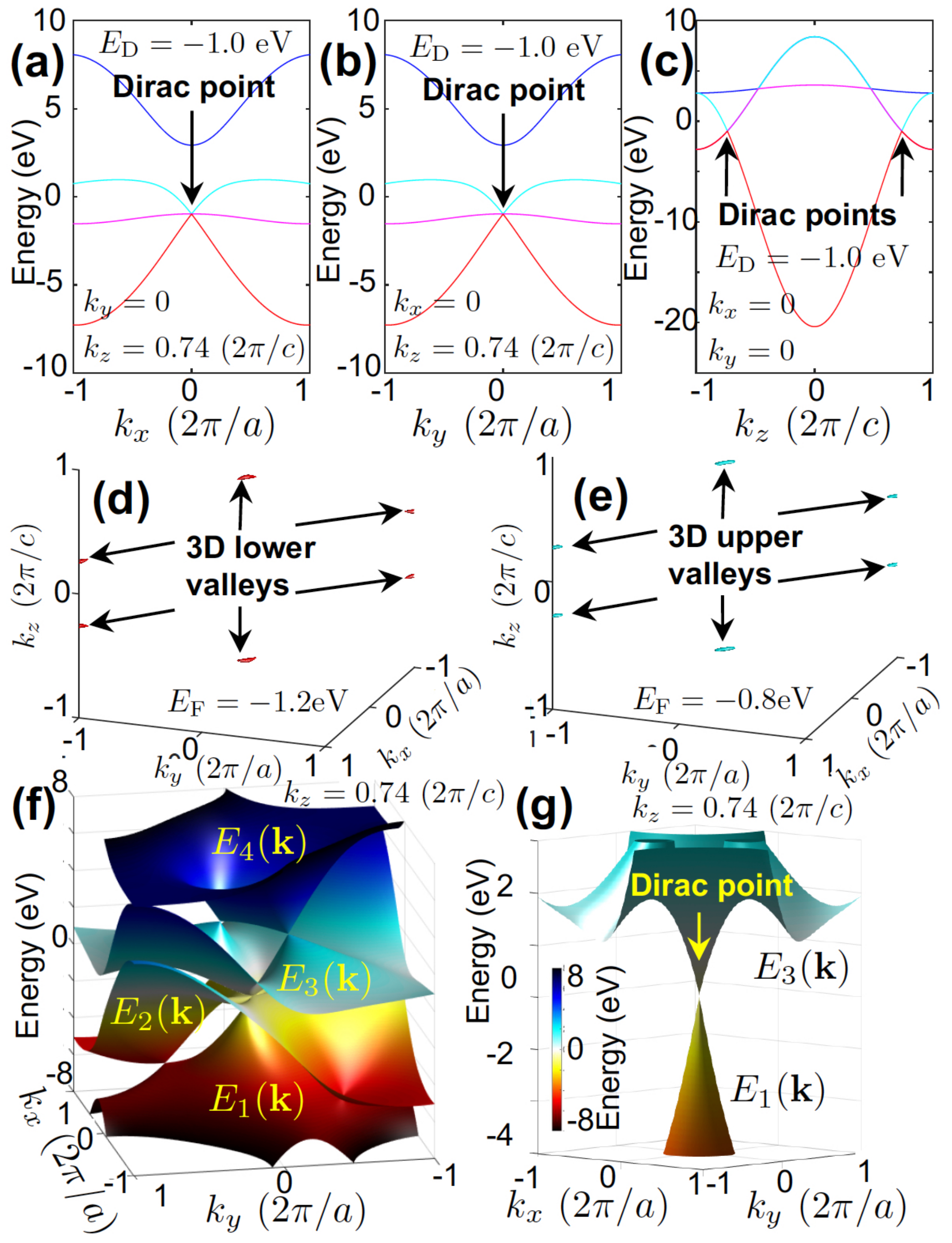}
\caption{
$3D$ Weyl Fermions of (2,2)-Hopfene.
Band structures at Dirac point ${\bf k}=(0,0,0.74)$ along (a) $x$, (b) $y$, and (c) $z$ directions.
Fermi surfaces of (d) lower and (e) upper valleys are also shown, proving the $3D$ feature.
(f) Energy band diagram in the $k_x$-$k_y$ plane at $k_z=0.74$, showing (a) linear dispersions of $E_3 ({\bf k})$ and $E_1 ({\bf k})$ with the flat band  $E_2 ({\bf k})$ in-between.
}
\end{center}
\end{figure}

For a stronger Hopf-link of $t_{\rm H}=1.5$ eV, we have calculated band structures by solving $\hat{H}$ numerically (Figs. 3-5). 
We confirmed $3D$ Weyl Fermions, which have linear dispersions along all directions (Figs. 3 (a)-(c)), appeared near the degenerate Dirac points with the energy of $E_{\rm D}=-1.0$ eV (Fig. 3).
We found the $3D$ Fermi surfaces, which should be spheroids, found at the Fermi energy $E_{\rm F}$ of -1.2 eV (Fig. 3 (d)) for lower valleys, and they disappeared at $E_{\rm D}$ as we increased  $E_{\rm F}$, while they appeared again at  $E_{\rm F}$ of -0.8 eV (Fig. 3 (e)) in the same locations of ${\bf k}$-space for higher valleys.
We also checked the linearity of the band at the Dirac point by plotting in both $k_x$-$k_y$ (Figs. 3 (f) and 3 (g)) and $k_x$-$k_z$ (Fig. 2 (b)) planes, so that the Weyl bands were formed between $E_3 ({\bf k})$ and $E_1 ({\bf k})$ .
It was also identified that the flat-band $E_2 ({\bf k})$ is penetrating between  $E_3 ({\bf k})$ and $E_1 ({\bf k})$, along  $k_x$ and $k_y$ directions, while $E_2 ({\bf k})$ is degenerate with $E_3 ({\bf k})$ along $k_z$ (Fig. 3 (c)).
These features are topologically equivalent to those expected at the weak coupling limit.

\begin{figure}[t]
\begin{center}
\includegraphics[width=8.6cm]{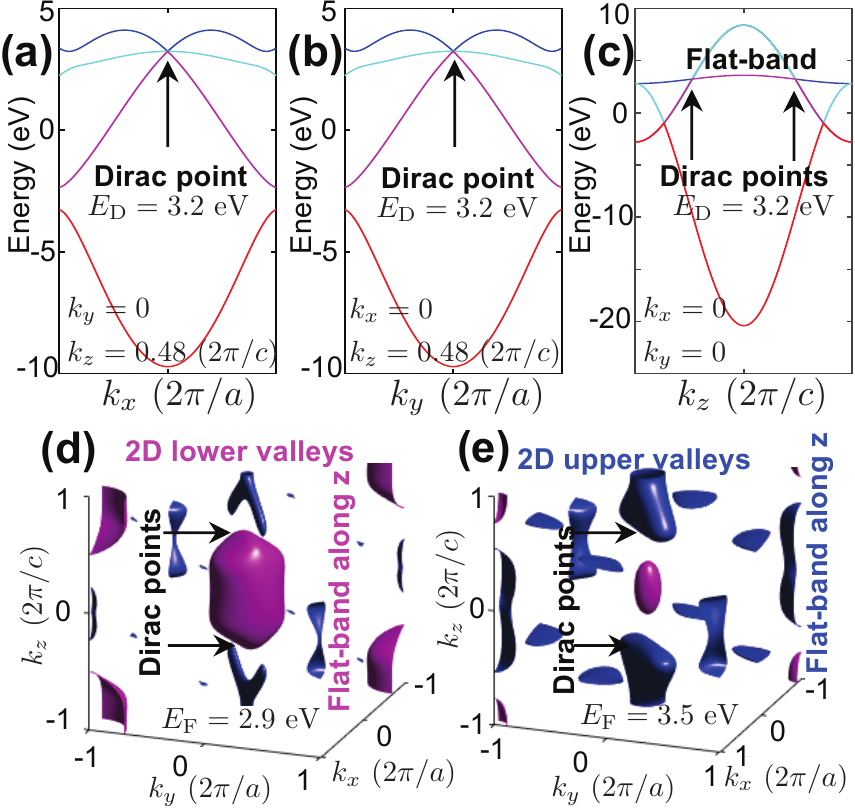}
\caption{
$2D$ Dirac Fermions of (2,2)-Hopfene.
Band structures at Dirac point ${\bf k}=(0,0,0.48)$  along (a) $x$, (b) $y$, and (c) $z$ directions.
Fermi surfaces of (d) lower and (e) upper valleys.
}
\end{center}
\end{figure}

At the higher energy of $E_{\rm D}=3.2$ eV, we found $2D$ Dirac Fermions at ${\bf k}=(0,0,0.48)$ (Fig. 4).
Near this point, the dispersions were linear along $k_x$ (Fig. 4 (a)) and $k_y$  (Fig. 4 (b)) , but it was almost flat along $k_z$  (Fig. 4 (c)).
This means that the confinement along $k_z$-direction was weak, so that the Fermi surfaces was spreading significantly along $k_z$ (Figs. 4 (d) and (e)).
Consequently, the cross sections of the Fermi surfaces along the $k_x$-$k_y$ plane were circular, leading to the $2D$ electronic characters.
These $2D$ Dirac Fermions are similar to those in a single Graphene sheet with regard to the $2D$ topological shape of the Fermi surface.
However, they also have dispersion along $k_z$ in Hopfene, so that they should be described as {\it quasi-}$2D$, strictly speaking, similar to correlated electronic materials such as cuprates\cite{Bednorz86}, in contrast to the monolayer of Graphene, where the perfect $2D$ confinement was achieved.

We also found the {\it quasi-}$1D$ Dirac Fermions, which were made of flat-band along the $k_x$-$k_y$ plane (Figs. 5 (a) and (b)), while they have linear dispersions along $k_z$ (Fig. 5 (c)).
In this case, the highest energy states in ${\bf k}$-space will become Fermi points along $k_z$ (Figs. 5 (d) and (e))
The electronic properties of these states will be {\it quasi-}$1D$.
We have 8 different $1D$ Dirac points in the plot of the Fermi surface of Figs. 5 (d) and (e).
Generations of these points would be linked to the original Dirac points and flat-bands.
If we had the Dirac points coming from Graphene or Hopf links, these points will survive upon the introduction of the adiabatic turn-on of $t_{\rm H}$, if $t_{\rm H}$ is sufficiently small.
Suppose we extend this linear dispersion from Dirac points to meet the flat-band, another Dirac points will be generated at the crossing points.
These generations should accompany both positive and negative topological charges with opposite chiralities\cite{Armitage18}, and thus the number of generated Dirac points would be even. 
In our Hopfene, we found huge amounts of Dirac points, characterised by different dimensionality.

\begin{figure}[h]
\begin{center}
\includegraphics[width=8.6cm]{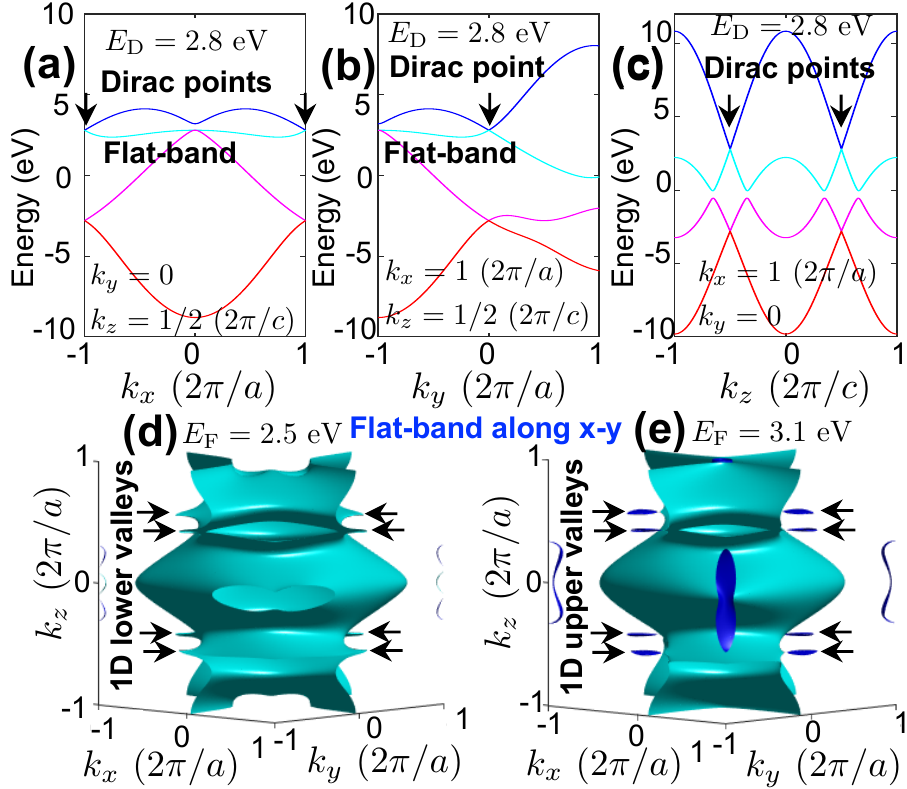}
\caption{
$1D$ Dirac Fermions of (2,2)-Hopfene.
Band structures at Dirac point ${\bf k}=(1,0,1/2)$ at $E_{\rm D}=2.8$ eV along (a) $x$, (b) $y$, and (c) $z$ directions.
Fermi points of (d) lower and (e) upper valleys.
}
\end{center}
\end{figure}

In the present letter, we have assumed that we can control Fermi level at our disposal.
In the real experiments, the Fermi level must be determined to ensure the charge neutrality.
In our calculation, $E_{\rm F}$ in non-doped (2,2)-Hopfene was 1.1 eV.
Due to the semi-metal structure of the electronic bands, it will not be possible to change the Fermi level by the application of the gate voltage in a metal-oxide-semiconductor structure.
However, the metal intercalation such as doping of alkali-metal would be possible\cite{Hebard91}, because of the huge available space along $z$ ($c$-axis) surrounded by Graphene sheets (Figs. 1 (a) and (c)).  
The flat-bands found in this letter would be favourable to possible observations of superconductivity\cite{Bardeen57} and/or ferromagnetism\cite{Auerbach94}, because of the increase in the density-of-states. 

In conclusion, electronic structures of the $3D$ Hopf-linked honeycomb lattices, named Hopfene, shows the distinctive topological features characterised by $3D$, $2D$, and $1D$ Weyl and Dirac Fermions.
The Fermi surfaces show the dimensional crossover upon changes of the Fermi level.
The impacts of topological structures on electrical properties are significant and the materials will be useful as platforms to examine various concepts of physics including magnetic monopoles, geometrical phases, and topological superconductors.

\section*{Acknowledgements}
This work is supported by EPSRC Manufacturing Fellowship (EP/M008975/1).
We would like to thank Prof. H. Mizuta, Dr. M. Muruganathan, Prof. Y. Oshima,  Prof. S. Matsui, Prof. S. Ogawa, Prof. S. Kurihara, and Prof. H. N. Rutt for stimulating discussions.
S.S also would like to thank JAIST for their hospitalities during his stay at the Center for Single Nanoscale Innovative Devices.

%

\end{document}